\documentclass[%
mathleft,%
]{an}
\usepackage{graphicx}
\usepackage{times}
\usepackage{fancyhdr}
\usepackage{amsmath}
\begin{document}

\Pagespan{1}{}
\Yearpublication{2011}%
\Yearsubmission{2011}%
\Month{1}%
\Volume{999}%
\Issue{92}%

\title{The {\em Fermi}-LAT view of young radio sources}

\author{F.\,D'Ammando\inst{1,2}\fnmsep\thanks{Corresponding author:
    \email{dammando@ira.inaf.it}} M.\,Orienti\inst{2}, M.\,Giroletti\inst{2}
  on behalf of the {\em Fermi} Large Area Telescope Collaboration
}
\titlerunning{The Fermi-LAT view of young radio sources}
\authorrunning{F.\,D'Ammando, M.\,Orienti, \and M.\,Giroletti}
\institute{Dip. di Fisica e Astronomia, Universit\`a di Bologna, Via Ranzani 1, I-40127 Bologna, Italy
\and  
INAF-Istituto di Radioastronomia, Via Gobetti 101, I-40129 Bologna, Italy}



\abstract{Compact Symmetric Objects (CSO) are considered to be the young version of Fanaroff-Riley type I and type II radio galaxies, with typical  sizes smaller than 1 kpc and ages of the order of a few thousand years. Before the launch of the {\em Fermi} satellite, young radio sources
  were predicted to emerge as a possible new $\gamma$-ray emitting population detectable by the Large Area Telescope (LAT). After more than 6 years of {\em Fermi} operation the question of young radio sources as $\gamma$-ray emitting objects still remains open. In this contribution we discuss candidate $\gamma$-ray emitting CSO and future perspective for detecting young radio sources with {\em Fermi}-LAT.}

\maketitle

\section{Introduction}

The discovery of emission in the $\gamma$-ray domain from many active galactic nuclei (AGN) by the EGRET telescope on board the {\em Compton Gamma-Ray Observatory} and the Che-renkov Telescopes was one of the most important breakthrough of high energy astrophysics in the last 20 years, leading to the identification of a new class of AGN: the blazar population (Punch et al. 1992, Hartman et al. 1999). The first 4 years of observations by the Large Area Telescope (LAT) on board the {\em Fermi Gamma-Ray Space} telescope confirmed that the extragalactic $\gamma$-ray sky is dominated by blazars and some radio galaxies (Acero et al. 2015). In addition, the discovery by {\em Fermi}-LAT of variable $\gamma$-ray emission from a few radio-loud narrow-line Seyfert 1 galaxies revealed the presence of a possible third class of AGN with relativistic jets (e.g., Abdo et al. 2009a, D'Ammando et al. 2015). Other classes of AGN may be able to emit up to the $\gamma$-ray energy range, and therefore to be detected by {\em Fermi}-LAT (e.g., the Circinus galaxy; Hayashida et al. 2013). In particular, young radio sources were predicted to constitute a relatively numerous class of extragalactic objects detectable by {\em Fermi}-LAT (Stawarz et al. 2008).

Powerful radio sources (L$_{1.4\rm\,GHz} >$ 10$^{25}$ W/Hz) are a small
fraction of the AGN, suggesting that the radio activity is a transient phase
in the life of these objects. The onset of radio emission is currently thought
to be related to mergers which provide fuel to the central AGN. In the
evolutionary scenario, the size of a radio source is strictly related to its
age. The discovery of the population of intrinsically compact and powerful
radio sources, known as compact symmetric objects (CSO), yielded an
improvement in the models proposed to link the various evolutionary stages of
the radio emission. CSO are characterized by linear sizes (LS) up to a few kpc, and a synchrotron spectrum
that turns over between hundreds of MHz and the GHz regime. CSO are considered
to represent an early stage in the radio source evolution and proposed as the
progenitors of the classical radio galaxies (Fanti et al. 1995, Readhead
et al. 1996, Snellen et al. 2000). Indeed, their radio morphology dominated by mini-lobes/hot spots resembles a scaled-down version of the edge-brightened Fanaroff-Riley type-II (FRII) galaxies (Fanaroff \& Riley 1974). The genuine youth of these objects is strongly supported by the determination of both the kinematic (e.g., Polatidis \& Conway 2003) and radiative (e.g., Murgia 2003) ages, which resulted to be in the range of 10$^{3}$-10$^{5}$ years, i.e. much shorter than those estimated in FRII galaxies (10$^{7-8}$ years; e.g., Parma et al. 2007). \\
Although young radio sources are preferentially studied in the radio band, the knowledge of their high-energy emission is crucial for providing
information on the most energetic processes associated with these sources, the actual region responsible for the high-energy emission, as well
as the structure of the newly born radio jets. In the next Section we briefly review the high-energy emission processes in young radio sources. The
presence of candidate young radio sources in the Third LAT AGN catalog is discussed in
Section 3. In Section 4 we present preliminary results of the $\gamma$-ray analysis of a
sample of bona-fide young radio sources, while the perspective for detecting
these objects with {\em Fermi}-LAT is discussed in Section 5. A brief summary
is presented in Section 6. 

\section{High-energy emission in young radio sources} 

The advent of the {\em Chandra} and {\em XMM-Newton} satellites revealed the X-ray emission of CSO (e.g., Siemiginowska et al. 2008, Tengstrand et
al. 2009). However, the extreme compactness of these sources and the insufficient spatial resolution of the current X-ray instruments could not locate the site of the X-ray
emission, and discriminate between the thermal contribution from the disc-corona system and the non-thermal emission from the extended structures
(e.g., Migliori et al. 2015). Indeed, a significant X-ray emission can be produced by jet, hot spots, and lobes as proved by the detection in FRII
galaxies (e.g., Hardcastle et al. 2002, Brunetti et al. 2002, Sambruna et al. 2004, Orienti et al. 2012).  For this reason $\gamma$-ray
emission from these sources may be detectable by {\em Fermi}-LAT. 

{\em Fermi}-LAT detected a flux of $\sim$10$^{-7}$ ph cm$^{-2}$ s$^{-1}$
from the lobes of the nearby ($z = 0.00183$, i.e. $d_L$ = 3.7 Mpc) radio
galaxy Centaurus A, corresponding to an apparent $\gamma$-ray luminosity of about
3$\times$10$^{41}$ erg s$^{-1}$. The $\gamma$-ray emission from the lobes, which
constitutes more than 50 per cent of the total source emission, is interpreted as
IC scattering of the cosmic microwave
background photons, with additional contribution at higher energies from the
infrared-to-optical extragalactic background light (Abdo et al. 2010).  If we consider a
young radio galaxy at redshift $z = 0.1$ ($d_L$ = 454.8 Mpc) with the same
apparent luminosity of the lobes of Centaurus A, this source should have a $\gamma$-ray flux of 1.2$\times$10$^{-14}$ ph
cm$^{-2}$ s$^{-1}$, far below the {\em Fermi}-LAT sensitivity limit.

However, given their compact size, CSO entirely reside within the innermost
region of the host galaxy, enshrouded by the dense and inhomogeneous
interstellar medium, which may be a rich source of UV/optical/IR seed photons. In the most compact radio sources, the radio lobes are only a few
parsecs from the AGN and their relativistic electrons can scatter the thermal
seed photons produced by both the accretion disc in UV and the torus in
infrared up to high energies (Stawarz et al. 2008). This mechanism is most
likely dominant in young radio sources optically associated with galaxies, where the projection effects should
be marginal. The high-energy luminosity strictly depends on different
parameters: source LS, jet power, UV/optical/IR photon density, and the
equipartition condition in the lobes. The most compact and powerful CSO with
energy equipartition in the lobes fulfilled and at a distance up to $\sim$ 1
GeV should be detectable in $\gamma$-rays. 

In CSO associated with quasars, the mechanism at the basis of the high-energy emission may be inverse Compton (IC) scattering off synchrotron seed photons produced by a different relativistic electron population. The main ingredient is the presence of a velocity gradient along the jet (e.g., Migliori et al. 2012). In this model the jet has two emitting regions  which are radiatively interacting: a fast blazar-like knot and a mildly relativistic knot outward. The central knot radiation is relativistically beamed due to the small viewing angle and illuminates the slower knot. The outer knot upscatters the synchrotron photons from the inner knot and radiates isotropically. 

Alternatively, the high-energy emission may be due to IC of the synchrotron photons by a dominant jet component, producing an emission that can be strongly beamed (e.g., Migliori et al. 2014). Seed photons for the IC mechanism can be the synchrotron emission of the knot (i.e., synchrotron self-Compton process) and the external radiation fields (the accretion disc and the dust surrounding the central engine; i.e., external Compton process). The main ingredients in this scenario are both the velocity of the emitting region, namely a jet knot, and its direction, approaching or moving away, relative to the source of the thermal/non-thermal photon seeds.

\begin{figure}
\centering
\includegraphics[width=8.0cm]{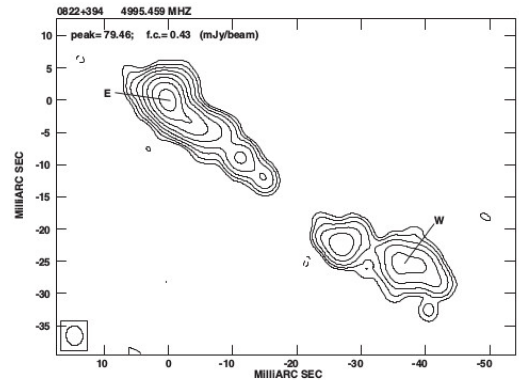}
\caption{VLBA image at 5.0 GHz of 4C $+$39.23B. Adapted from Orienti et al. (2004).}
\label{4c39}
\end{figure}

\section{Young Radio Sources in the Third LAT AGN Catalogue}

By considering the possible non-thermal high-energy emission from lobes in galaxies and from jets in quasars, CSO are expected to be $\gamma$-ray emitting sources and it is interesting to check if some of them are included in the last {\em Fermi}-LAT catalogue. The third catalog of AGN detected in $\gamma$-rays by {\em Fermi}-LAT (3LAC; Ackermann et al. 2015) is based on four years of LAT data collected between 2008 August 4 and 2012 July 31. The 3LAC includes 1591 AGN located at high Galactic latitudes ($|b|$ $>$ 10$^{\circ}$). Most of the objects (98\%) are blazars, but a few of them may be a young radio source. In particular, three objects are tentatively classified as a compact steep spectrum (CSS) object: 4C $+$39.23B, 3C 286, and 3C 380. 

\begin{figure}
\centering
\includegraphics[width=7.0cm]{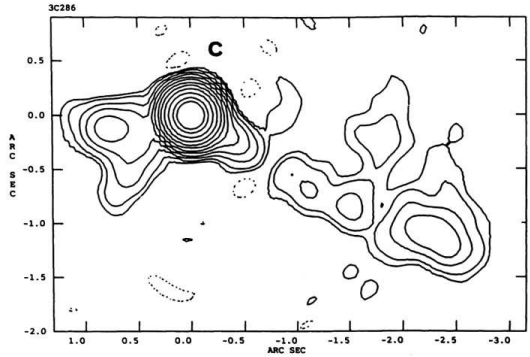}
\caption{MERLIN image at 1.6 GHz of 3C 286. Adapted from Akujor et al. (1994).}
\label{3C286}
\end{figure}

4C $+$39.23B is a compact  radio source with a double morphology dominated by steep-spectrum lobes, with a linear size of $\sim$200 pc (Fig.~\ref{4c39}; Orienti et al. 2004). In the 3LAC the source may be associated with the $\gamma$-ray source 3FGL J0824.9$+$3916. However, the non-unique association of the $\gamma$-ray source (3FGL J0824.9$+$3916 is also associated with the flat spectrum radio quasar 4C $+$39.23) makes its CSO counterpart unlikely.

3C 286 and 3C 380 are bright steep spectrum radio quasars
(SSRQ) at redshift $z$ = 0.849 and $z$ = 0.692, respectively. 3C 286,
associated with the $\gamma$-ray source 3FGL J1330.5$+$3023, has the morphology of a medium-sized
symmetric object (MSO) (Fig.~\ref{3C286}). Cotton et al. (1997) suggested that 3C 286 may be a core-jet source in which the
relativistic core is beamed away from our line of sight and it is not
detected, and what we see is the jet that curves through our line of sight. 

\begin{figure*}
\begin{center}
\rotatebox{0}{\resizebox{!}{60mm}{\includegraphics{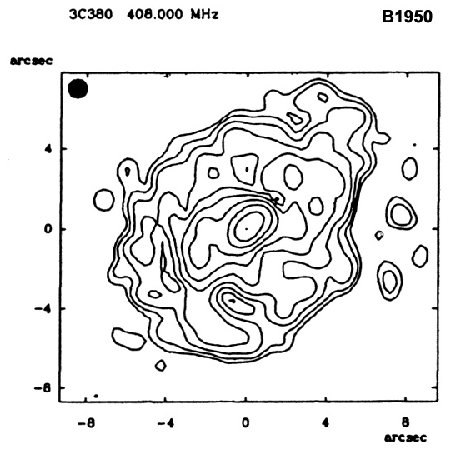}}}
\hspace{2cm}
\rotatebox{0}{\resizebox{!}{60mm}{\includegraphics{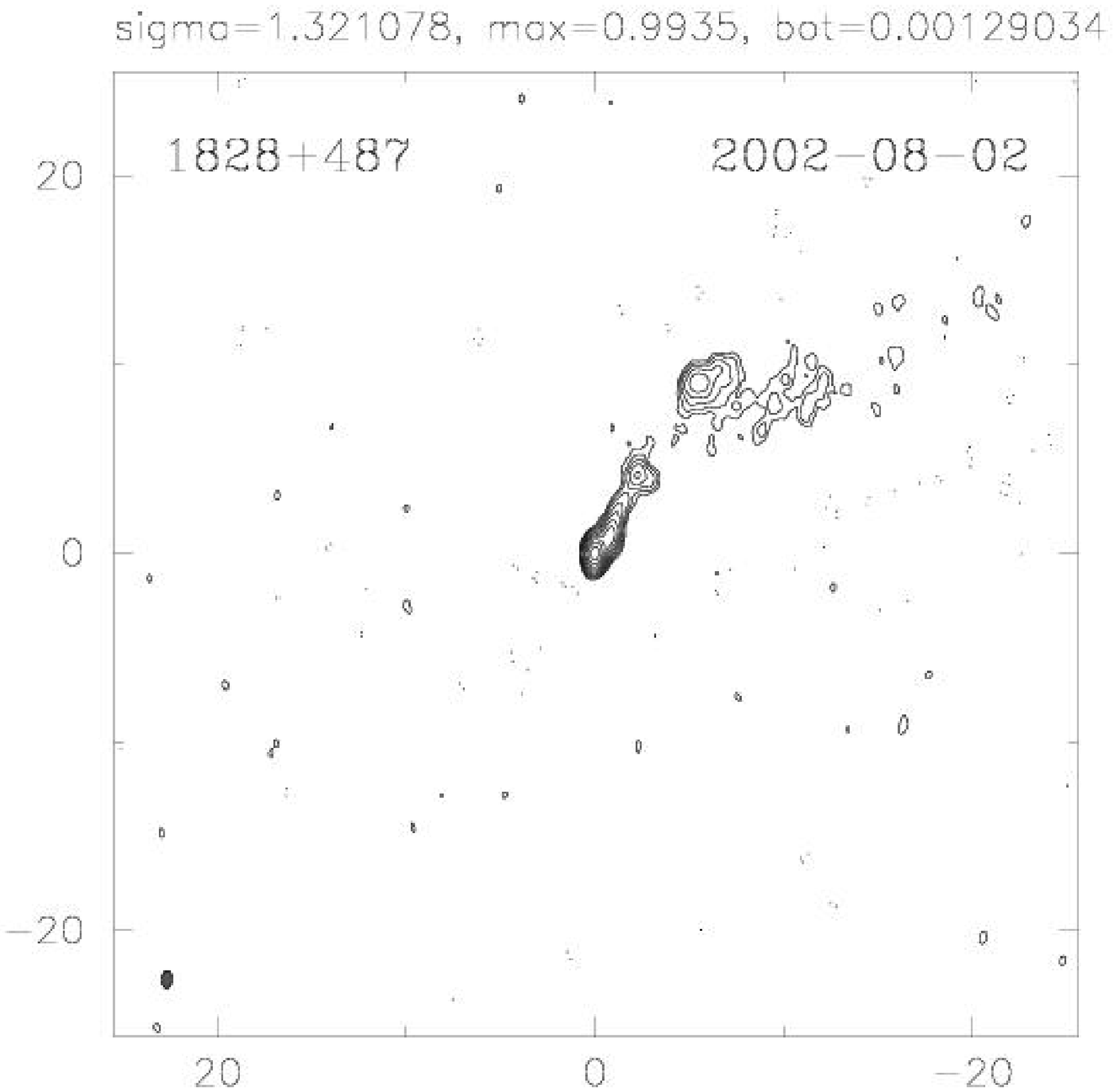}}}
\caption{{\it Left panel}: MERLIN image at 408 MHz of 3C 380. Adapted from Reid et al. (1995). {\it Right panel}: VLBA image at 15 GHz of 3C 380. Adapted from Lister \& Homan (2005).}
\label{3C380}
\end{center}
\end{figure*}

3C 380, associated with the $\gamma$-ray source 3FGL \\ J1829.6$+$4844, was part
of the sample of powerful CSS objects selected by Fanti et al. (1990). However, its
complex radio structure extending on a total linear size of $\sim 70$ kpc
(largely exceeding the optical host galaxy) (Fig.~\ref{3C380}) and the detection of superluminal
motion along the jet (Polatidis \& Wilkinson 1998) suggest that 3C 380 is a
classical SSRQ, rather than a young CSO.

In addition, three other $\gamma$-ray sources included in the 3LAC are claimed as possible CSO quasars: 4C
$+$55.17, PMN J1603$-$4904, and PKS 1413$+$135. 
 
4C $+$55.17 ($z$ = 0.896) first appeared as a $\gamma$-ray source during the
EGRET era (as 3EG J0952+5501, Hartmann et al. 1999, and EGR J0957+5513, Casandjian \& Grenier 2008). The tentative association, due to poor EGRET localization, was confirmed by {\em Fermi}-LAT after the
first three months of observations (Abdo et al. 2009b). The VLBI morphology at
5 GHz reveals two distinct emission regions, which may resemble compact hot
spots and lobes. In this scenario the kpc-scale emission observed with the VLA
might be interpreted as a remnant of previous jet activity. Based on the radio
morphology, Rossetti et al. (2005) first suggested a classification as MSO for
this object. The source showed no evidence of blazar flaring activity at any
wavelength, nor any significant evidence of long-term variability from radio
to $\gamma$-rays. Moreover, the low brightness temperature and the hard
$\gamma$-ray spectrum observed by {\em Fermi}-LAT is unlikely for a flat
spectrum radio quasar (FSRQ). However, the presence of broad optical emission
lines in its spectrum and high optical/UV core luminosity, together with its
high $\gamma$-ray luminosity (L$_{\gamma}$ $\sim$ 10$^{47}$ erg s$^{-1}$) have
in turn led to the common classification of 4C $+$55.17 as a FSRQ. The
spectral energy distribution (SED) of the source is modelled by both the blazar and CSS scenario (McConville et al. 2011), leaving the debate on the nature of this object still open. 

PMN J1603$-$4904 has been classified as a low-synchrotron-peaked  BL  Lac
object in the Second Fermi LAT Source Catalog (Nolan et al. 2012). The source
is bright and variable in $\gamma$-rays, in agreement with this classification. However, TANAMI VLBI observations showed a resolved and
symmetric brightness distribution on milliarcsecond scales (M\"{u}ller et al. 2014). The brightest
component in the center shows a flat spectrum and the highest brightness
temperature. This result is in contrast to the typical BL Lac structure, where
the optically thick core component is expected to be at the bright end of a one-sided jet structure. Moreover, the SED shows a high Compton
dominance and a peculiar IR excess
with T = 1600 K not expected in BL Lac objects. The detection of a redshifted Iron line in the X-ray spectra
collected by {\em XMM-Newton} and {\em Suzaku} suggests that PMN J1603$-$4904
is observed at a larger viewing angle than that expected for blazars, challenging
its classification as a BL Lac object (M\"{u}ller et al. 2015).

PKS 1413$+$135 was initially classified as a BL Lac object because of its
optical properties (Beichman et al. 1981). However, based on its radio
properties Peck \& Taylor (2000) included PKS 1413$+$135  in the sample of CSO
in the northern sky (COINS). Its morphology consists of a two-sided structure
with a jet, a bright flat-spectrum core, and the \\ counter-jet (Gugliucci et
al. 2005), which is similar to the radio structure of PMN J1603$-$4904. A clear difference in the
length and brightness of the jet and counter-jet was observed for PKS
1413$+$135, suggesting an orientation close to the line of sight, and therefore
a classification as a CSO quasar with superluminal jet motion (Gugliucci et al. 2005).

\section{Searching for $\gamma$-ray emitting young radio sources}

We are investigating the $\gamma$-ray emission in a sample of 51 bona-fide
young radio sources selected by Orienti \& Dallacasa (2014). These objects are
associated with galaxies or SSRQ to avoid severe boosting effects, with a known redshift, and a core detection. 

We analysed 6 years of LAT data (2008 August 4 -- 2014 August 4) in the 0.1--10 GeV energy range. Details about the LAT are given by Atwood et
al. (2009). The analysis was performed with the \texttt{ScienceTools} software package version v10r0p5\footnote{http://fermi.gsfc.nasa.gov/ssc/data/analysis/}. The LAT data were extracted within a 10$^{\circ}$ region of interest  centred at the location of the target. Only events belonging to the ``Source'' class were used. In addition, a
cut on the zenith angle ($<$100$^{\circ}$) was applied to reduce
contamination from the Earth limb $\gamma$-rays, which are produced by cosmic
rays interacting with the upper atmosphere. The spectral analysis was
performed with the instrument response functions P7REP\_SOURCE\_V15 using an
unbinned maximum likelihood method implemented in the Science tool
\texttt{gtlike}. Isotropic (`iso\_source\_v05.txt') and Galactic diffuse
emission (`gll\_iem\_v05\_rev1.fit') components were used to model the
background\footnote{http://fermi.gsfc.nasa.gov/ssc/data/access/lat/\\BackgroundModels.html}. The
normalization of both components was allowed to vary freely during the
spectral fitting. 

We evaluated the significance of the $\gamma$-ray signal from the source by
means of a maximum-likelihood test statistic (TS) that results in TS = 2$\times$(log$L_1$ $-$ log$L_0$), where
$L$ is the likelihood of the data given the model with ($L_1$) or without ($L_0$) a point source at the position of the target (e.g., Mattox et al. 1996). The source model used in
\texttt{gtlike} includes all the point sources from the 3FGL catalogue that
fall within $15^{\circ}$ from the target source. The fitting procedure has been performed with the
sources within 10$^{\circ}$ from the target source included with the
normalization factors and the photon indices left as free parameters. For the
sources located between 10$^{\circ}$ and 15$^{\circ}$ from our target we kept the
normalization and the photon index fixed to the values of the 3FGL catalogue. We removed all sources with TS $<$ 25. 
A second maximum likelihood analysis was performed on the updated source model. 

\noindent No significant detection (TS $>$ 25) over 6 years were obtained for 
the sources analyzed so far, with a 2$\sigma$ upper limit ranging between (0.7--11.5)$\times$10$^{-9}$ ph cm$^{-2}$ s$^{-1}$.

The event selection developed for the LAT has been periodically updated to reflect the constantly improving knowledge of the
detector and the environment in which it operates. A radical revision of the
entire event-level analysis has been performed with the new Pass 8 data. The
main improvements are an increased effective area, a better point-spread
function, and a better understanding of the systematic uncertainties. The new
event reconstruction allows us to extend the energy reached by the LAT below 100 MeV and
above 1 TeV (Atwood et al. 2013). In particular, analysis of Pass 8 data below 100 MeV
may be important for a class of sources for which the photon index is expected
to be soft, and thus mainly emitting at low energies. We plan to analyze Pass 8 data of the
sample of sources studied here.

\section{Perspective for $\gamma$-ray detection}

\begin{figure}
\centering
\includegraphics[width=8.0cm]{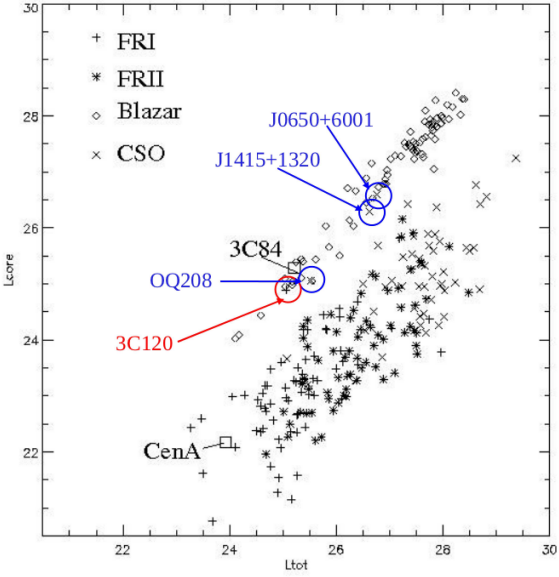}
\caption{Total luminosity (x-axis) vs core luminosity (y-axis) of a sample of FRI (+ signs) and FRII (asterisks) radio galaxies from Giovannini et
  al. (2001), blazars in the LBAS (diamonds), and CSO (x simbols). Adapted from Orienti et al. (2011).}
\label{Lcore}
\end{figure}

A correlation between the core luminosity and the total luminosity for FRI
and FRII radio galaxies was found in Giovannini et al. (1998). In Orienti et
al. (2011) the contribution of the core component to the total
luminosity was investigated for CSO (Fig.~\ref{Lcore}). This class of objects seems to extend the
correlation found for FRI and FRII to higher luminosities. This is expected since CSO are mainly detected at high redshift (0.4
$ < z <$ 2), with only a few objects at $z$ $\sim 0.1$. Interestingly, in the
core luminosity vs total luminosity plot 3 CSO lie in the region occupied by blazars in the {\em Fermi} LAT Bright AGN Sample (LBAS; Abdo et al. 2009c): OQ 208, J0650$+$6001, and PKS 1413$+$135, making these CSO as good
candidates for high-energy emission. As shown in Section 3, PKS 1413$+$135 is
associated with a $\gamma$-ray source in the 3LAC catalogue.

As discussed in Section 2, a possible mechanism for
producing high energy emission may be the IC of thermal
UV photons by the lobes' relativistic electrons. Under the
assumption of equipartition and assuming that the jet luminosity is
L$_{\rm j}$ = 10$\times$L$_{\rm tot}$, 
and the luminosity provided by the UV photons is 
L$_{\rm UV}$ = 10$^{46}$ erg s$^{-1}$, 
we computed the expected luminosity at 1 GeV for OQ\,208 and J0650+6001 
using the formula from Stawarz et al. (2008):

\begin{equation} 
\begin{split} 
{[\varepsilon L_{\varepsilon}]_{\rm IC/UV} \over 10^{42} \, {\rm
    erg/s}} \sim 2 \, \left({\eta_{\rm e} \over \eta_{\rm B}}\right)
\left({L_{\rm j} \over 10^{45} \, {\rm erg/s}}\right)^{1/2} \left({LS
  \over 100 \, {\rm pc}}\right)^{-1} \times \\
\left({L_{\rm UV} \over 10^{46} \, {\rm erg/s}}\right) \left({\varepsilon \over 1 \, {\rm GeV}}\right)^{-0.25}
\label{lum}
\end{split}
\end{equation}
\noindent
or the appropriate IC/UV energy flux:
\begin{equation}
{[\varepsilon S_{\varepsilon}]_{\rm IC/UV} \over 10^{-12} \, {\rm erg / cm^{2} / s}} \sim 1.6 \times \left({[\varepsilon L_{\varepsilon}]_{\rm IC/UV} \over 10^{42} \, {\rm erg/s}}\right) \left({d_{\rm L} \over 100 \, {\rm Mpc}}\right)^{-2} ,
\label{flux}
\end{equation}
\noindent
where $d_{\rm L}$ is the luminosity distance to the source, LS is the source
linear size, $\eta_{\rm e}$/$\eta_{\rm B}$ is the ratio between the
particle energy and the magnetic field energy.\\
By means of Equations \ref{lum} and \ref{flux}, we computed the
expected 1-GeV luminosity and flux density for OQ\,208 and \\ J0650+6001,
which turn out to be: 

\begin{itemize}
\item OQ\,208: L$_{1\rm\,GeV}$ = 2.8 $\times$ 10$^{44}$ erg s$^{-1}$, S$_{1\rm\,GeV}$ = 3.8 $\times$
  10$^{-11}$ erg cm$^{-2}$ s$^{-1}$; \\

\item J0650+6001: L$_{1\rm\,GeV}$ = 5 $\times$ 10$^{44}$ erg s$^{-1}$, S$_{1\rm\,GeV}$ = 1.2 $\times$
  10$^{-12}$ erg cm$^{-2}$ s$^{-1}$. \\
\end{itemize}

\noindent Assuming standard parameters as above, OQ\,208, that is one of the
closest CSO ($z = 0.076$), should have been detected with the sensitivity obtained at 1 GeV
in 4 years of observations by {\em Fermi}-LAT for sources outside the Galactic
plane, i.e.~2$\times$10$^{-12}$ erg cm$^{-2}$ s$^{-1}$ at 5$\sigma$. The non-detection of the source indicates that the parameters used in the model are too extreme, setting upper limits to the jet power and the amount of UV photons. 
The high redshift ($z > 0.4$) typical of the majority of young radio sources makes these objects even more difficult to detect.
However, in the case of CSO associated with SSRQ, like J0650+6001, where also moderate boosting effects should be present, we may
consider an additional contribution of IC made by relativistic electrons from
the jet enhanced by boosting effects (e.g., Ghisellini et al. 2005). 

\section{Conclusions}

The young radio sources as a population of $\gamma$-ray emitting objects remain
elusive up to now. The nature of some
$\gamma$-ray sources supposed to be CSO and detected by {\em Fermi}-LAT during the first 4 years of
operation has to be confirmed. In any case, the number of CSO detected in
$\gamma$-rays is significantly lower than what was expected from theoretical models. This may be due to different reasons: an overestimate of the jet
luminosity or UV photon energy density in these objects, a strong intrinsic
$\gamma$-$\gamma$ opacity, efficient emission mechanisms mainly below 100
MeV.

We investigated the $\gamma$-ray emission from a complete sample of bona-fide
young radio sources using 6 years of {\em Fermi} data, and we have not detected any
source at high significance so far. The detection of bona-fide young radio
sources in $\gamma$-rays would provide the confirmation
that the non-thermal emission is dominating the high-energy emission in these
objects. Moreover, $\gamma$-ray emission from the jet may allow us to shed a
light on the jet dynamics during the initial stages of the activity of these
sources, whether there is a single velocity or a more complex structure.

\acknowledgements  
The \textit{Fermi}-LAT Collaboration acknowledges support for LAT development,
operation and data analysis from NASA and DOE (United States), CEA/Irfu and
IN2P3/CNRS (France), ASI and INFN (Italy), MEXT, KEK, and JAXA (Japan), and
the K.A.~Wallenberg Foundation, the Swedish Research Council and the National
Space Board (Sweden). Science analysis support in the operations phase from
INAF (Italy) and CNES (France) is also gratefully acknowledged. 

Part of this work was done with the contribution of the Italian Ministry of Foreign Affairs
and research for the collaboration project between Italy and Japan. 

\end{document}